\documentclass[10pt,conference]{IEEEtran}
\IEEEoverridecommandlockouts
\usepackage{cite}
\usepackage{amsmath,amssymb,amsfonts}
\usepackage{algorithmic}
\usepackage{graphicx}
\usepackage{textcomp}
\usepackage{xcolor}

\usepackage{pifont}
\usepackage{multirow}
\def\BibTeX{{\rm B\kern-.05em{\sc i\kern-.025em b}\kern-.08em
    T\kern-.1667em\lower.7ex\hbox{E}\kern-.125emX}}
\begin{document}

\title{A Novel Spatial-Temporal Variational Quantum Circuit to Enable Deep Learning on NISQ Devices}


\author{\IEEEauthorblockN{
Jinyang Li\textsuperscript{\dag,\pounds},
Zhepeng Wang\textsuperscript{\dag,\pounds}, 
Zhirui Hu\textsuperscript{\dag,\pounds}, 
Prasanna Date\textsuperscript{\S},
Ang Li\textsuperscript{\ddag},
Weiwen Jiang\textsuperscript{\dag,\pounds}}

\IEEEauthorblockA{\textsuperscript{\dag}George Mason University, Department of Electrical and Computer Engineering, VA, USA.\\
\textsuperscript{\pounds}George Mason University, Quantum Science and Engineering Center, VA, USA.\\
\textsuperscript{\S}Oak Ridge National Laboratory, TN, USA.\\
\textsuperscript{\ddag}Pacific Northwest National Laboratory, WA, USA.\\
\{jli56, wjiang8\}@gmu.edu
\vspace{-0.15in}}
}

\maketitle

\begin{abstract}
Quantum computing presents a promising approach for machine learning with its capability for extremely parallel computation in high-dimension through superposition and entanglement. Despite its potential, existing quantum learning algorithms, such as Variational Quantum Circuits (VQCs), face challenges in handling more complex datasets, particularly those that are not linearly separable. 
What's more, it encounters the deployability issue, making the learning models suffer a drastic accuracy drop after deploying them to the actual quantum devices.
To overcome these limitations, this paper proposes a novel spatial-temporal design, namely ``ST-VQC'', to integrate non-linearity in quantum learning and improve the robustness of the learning model to noise.
Specifically, ST-VQC can extract spatial features via a novel block-based encoding quantum sub-circuit coupled with a layer-wise computation quantum sub-circuit to enable temporal-wise deep learning.
Additionally, a SWAP-Free physical circuit design is devised to improve robustness.
These designs bring a number of hyperparameters.
After a systematic analysis of the design space for each design component, an automated optimization framework is proposed to generate the ST-VQC quantum circuit.
The proposed ST-VQC has been evaluated on two IBM quantum processors, \textit{ibm\_cairo} with 27 qubits and \textit{ibmq\_lima} with 7 qubits to assess its effectiveness. The results of the evaluation on the standard dataset for binary classification show that ST-VQC can achieve over 30\% accuracy improvement compared with existing VQCs on actual quantum computers. Moreover, on a non-linear synthetic dataset, the ST-VQC outperforms a linear classifier by 27.9\%, while the linear classifier using classical computing outperforms the existing VQC by 15.58\%.
\end{abstract}

\section{introduction}
Quantum computing is rapidly growing and evolving. With the emergence of real-world quantum computers from companies like IBM, IonQ, and Quantinuum, the potential for solving complex problems that were previously intractable for traditional computers has become a reality. A promising application of quantum computing is in the realm of machine learning, known as quantum learning \cite{jiang2021co, wang2021exploration, jiang2021machine, stein2022quclassi, jing2022rgb, zeng2022multi, tacchino2020quantum, yan2020nonlinear, schuld2015introduction, biamonte2017quantum, ciliberto2018quantum,christensen2020fchl,mcclean2018barren, killoran2019continuous,krisnanda2021creating,wu2021scrambling,gratsea2022exploring, hu2022design, hu2022quantum, wang2023qumos, liang2021can}. These algorithms take advantage of the unique features of quantum computing, such as superposition and entanglement, to perform highly parallel and efficient computations. The ultimate goal of quantum learning is to develop algorithms that can tackle learning problems that are too challenging for traditional computers, including large-scale optimization problems, deep learning, and data classification.

The existing quantum learning algorithm, known as Variational Quantum Circuits (VQC), exhibits limitations in terms of the expressibility of the learning models. Specifically, it is restricted to performing linear operations and is deficient in its capacity to handle datasets that possess non-linear separability.
The absence of non-linearity in VQCs renders them unsuitable for deployment as deep learning models, as non-linearity constitutes a critical element in enabling models to capture complex input-output relationships. To surmount this constraint, novel techniques must be developed to integrate non-linearity into quantum learning circuits, thereby enhancing their utility for deep learning applications.

The design of computation circuits for quantum learning confronts several challenges. First, the design functions like a fully connected layer, thereby disregarding the spatial attributes of input data. This implies that it is unable to discern relationships between data points in a specific spatial configuration. This deficiency in spatial feature extraction can undermine the accuracy of the model. Second, the manual design of computation circuits does not consider the compilation process, which may necessitate a substantial number of SWAP gates. This can result in more errors due to noise, ultimately contributing to a significant reduction in accuracy.
To address these drawbacks, it is necessary to develop new techniques that can take into account the spatial features of the input data and optimize the circuit design for specific hardware architectures to reduce noise and improve accuracy.

As mentioned before, non-linearity is the most important part of quantum machine learning given that it can find the non-linear mapping relationship between input and output, which leads to the main question in this paper: How can we bring and enhance the non-linearity in VQC? 
This is a non-trivial and open question for quantum learning.
The challenges lie in the following two aspects:
(1) The unitary operations of quantum computation conduct the learning transformation from input to output;
(2) Although the measurement for qubits can introduce non-linearity, it has high costs in terms of both fidelity and latency.


In this paper, we tackle this hard nut from another angle: we propose a system-level automatic design framework to design a novel spatial-temporal VQC, namely ST-VQC, which enables deep learning on quantum processors. Here, the spatial-temporal refers to the consideration of extracting spatial information through a novel block-based encoding quantum sub-circuit, and the layer-wise design of computation sub-circuits that allow gradual entanglement of qubits to enable deep learning. There are two core design blocks in the framework: ST-Encoder and ST-Processor.
In ST-Encoder, we propose a nonlinear data encoder to integrate non-linearity into the forward propagation process of quantum neural networks and a spatial data encoder to improve the fidelity of encoding. In ST-Processor, we propose a logical circuit design to capture the local spatial features and equip a cascaded structure to enable non-linearity.
What's more, a 
SWAP-free physical circuit design is developed to avoid involving additional SWAP gates and improves the fidelity of computation. 
After a throughout analysis of the system, we formulate the design space for each component. We then develop a system-level optimization tool that can automatically generate ST-VQC quantum circuits.



The main contributions of this paper are as follows.

\begin{itemize}
    \item A novel data encoder, namely ST-Encoder, is proposed to overcome the lack of non-linearity in existing quantum neural networks and to make trade-offs between qubits' number and fidelity.
    \item A spatial- and compilation-aware data processor, namely ST-Processor, is developed  to extract the spatial features of input data and gradually integrate the non-linearity by a cascaded structure. Additionally, we bring the compilation information in the design of logical quantum circuits, which can significantly reduce the number of SWAP gates for improving robustness against noise.    
    \item A reinforcement learning-based optimization algorithm is further proposed to automatically identify the best hyperparameters in each design component of ST-Encoder and ST-Processor, which can find high-quality solutions while satisfying the resource limitations.
    
\end{itemize}

The proposed ST-VQC has been evaluated on two IBM quantum processors, \textit{ibm\_cairo} with 27 qubits and \textit{ibmq\_lima} with 7 qubits, to assess its effectiveness. Experimental results on MNIST-2 showed that while the existing VQC achieved over 90\% accuracy in simulation, its accuracy dropped to approximately 50\% when deployed on real quantum computers, rendering it ineffective.
In contrast, the proposed framework demonstrated the ability to find a quantum learning circuit with an accuracy of 90\% on real quantum computers. Furthermore, the results on non-linearly separable synthetic datasets showed that the proposed ST-VQC outperformed a linear classifier by 27.9\% of accuracy, while even the linear classifier on classical computing outperformed the existing VQC by 15.58\% of accuracy.


The remainder of the paper is organized as follows. 
Section \MakeUppercase{\romannumeral 2} provides the background and preliminary. Section \MakeUppercase{\romannumeral 3} discusses the observations and challenges in existing quantum learning. Section \MakeUppercase{\romannumeral 4} introduces the design of the proposed framework, while the system-level optimization is shown in Section \MakeUppercase{\romannumeral 5}. Experimental results and concluding remarks are given in Sections \MakeUppercase{\romannumeral 6} and \MakeUppercase{\romannumeral 7}.
\section{background and preliminary}



\subsection{Quantum Learning}

\begin{figure}[t]
    \centering
    \includegraphics[width=0.8\columnwidth]{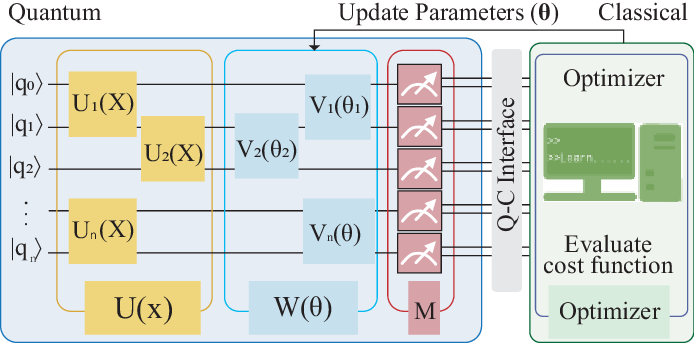}
    \caption{A general quantum learning framework}
    \label{fig:quantum-classcial}
\end{figure}

Variational quantum circuit (VQC) \cite{altaisky2001quantum,ezhov2000quantum,yan2020nonlinear}, also known as quantum neural network (QNN), is a quantum circuit with 
parameterized quantum gates, where the parameter in each gate is trainable. 
The blue box on the left-hand side in Figure~\ref{fig:quantum-classcial} illustrates a general design of a VQC, which mainly consists of three parts \cite{chen2020variational}: (1) Input Encoder $U(x)$: quantum gates with non-adaptable parameters for encoding classical data to quantum domain; Specifically, There are two commonly used encoding methods, which are amplitude encoding and angle encoding. Amplitude encoding is to convert the input values to the amplitudes of states $\mathbf{x}$ by L2-normalization to satisfy $\sum_{i=0}^{2^n}|{x}_{i}|^2=1$, while angle encoding is to put the input data into the parameter of rotation gates on each qubit sequentially.  (2) Trainable Quantum layers $W(\theta)$: quantum gates with adaptable parameters $\theta$; (3) Measurement $M$:  the measurement operations to project the quantum output into the classical domain, whose output will be used for prediction. 

Figure \ref{fig:quantum-classcial} shows a typical quantum learning framework, where the core component is a parameterized quantum circuit (a.k.a., variational quantum circuit or ansatz circuit).
Like classical machine learning, there are two steps in quantum learning: forward propagation and backward propagation.
The forward propagation only involves quantum computing, where data will be encoded, processed, and measured in the quantum circuit.
The backward propagation further involves classical computing to calculate the loss according to the measurement results and label, which will then be used to update the parameters in VQC.
The optimization framework is to find the optimal parameters in quantum circuits so that the cost function of a certain task can be minimized. Both gradient-based \cite{sweke2020stochastic} and gradient-free \cite{zhu2019training} optimization algorithms can be used to update parameters.

\subsection{Quantum Compilation}
\begin{figure}[t]
\centering
\includegraphics[width=1\linewidth]{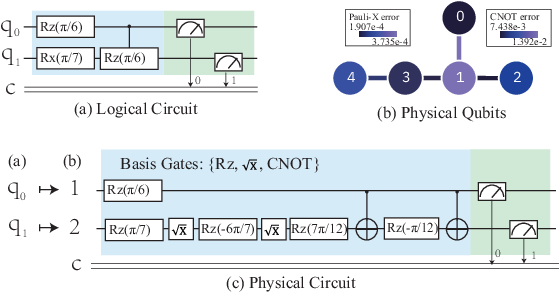}
\caption{Background of compilation: (a) logical quantum circuit with two qubits $q_0$ and $q_1$; (b) the topology of physical qubits with noise; (c) the physical circuit after mapping qubits from (a) to (b) and decompose quantum gates using basic gates.}
\label{fig:vqc_comp}
\end{figure}

The current VQC design~\cite{chen2020variational} is usually constructed with logical gates, called logical circuits, which cannot be deployed on real quantum computers directly. The logical circuit needs to be compiled into a physical circuit at first, in order to meet the hardware constraints of the real quantum computer and then be deployed onto it. The details of the compilation are shown in Figure~\ref{fig:vqc_comp}. 

More specifically, there are three steps ~\cite{venturelli2019quantum}: (1) Decompose the logical gates into physical basic gates supported by the real quantum computer; (2) Map the logical qubits to physical qubits, which produces a physical circuit; (3) Optimize the generated physical circuits for the shorter circuit depth.
After the compilation, the physical gates on the physical circuit are executed following the order of directed acyclic graph (DAG) \cite{liang2021can} of the circuit. 

\section{challenge in quantum learning}
\subsection{Lack of Non-Linearity in Forward Propagation of Existing Quantum Neural Networks}
\begin{figure}[t]
    \centering
    \includegraphics[width=1\linewidth]{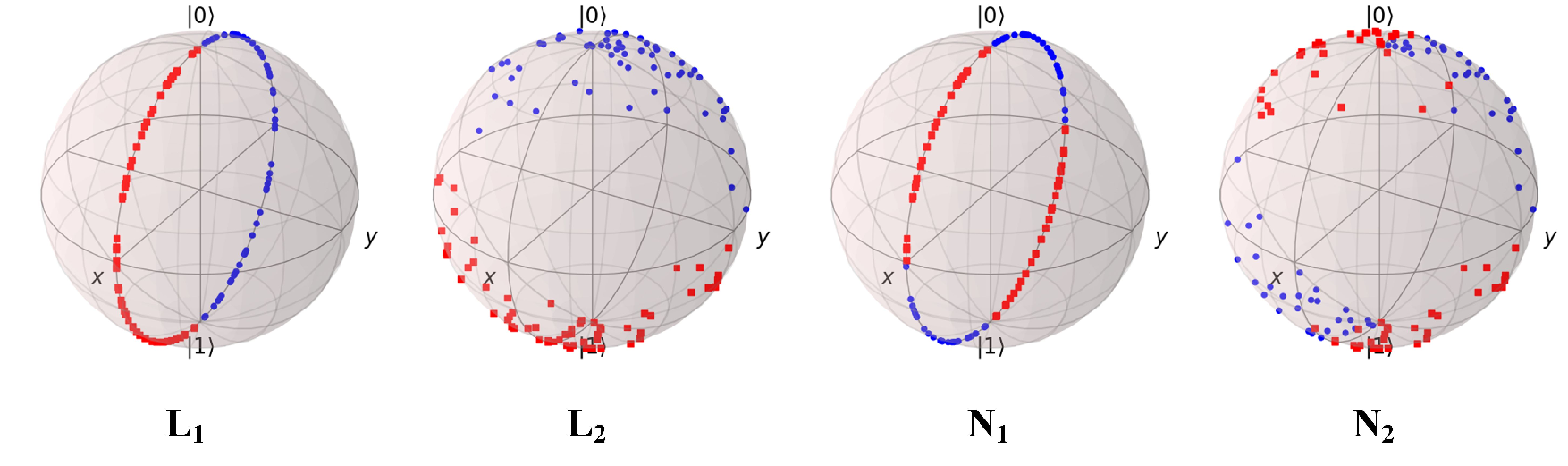}
    \caption{Four datasets used to evaluate existing quantum circuit models. L1 and L2 refer to linearly separable datasets. N1 and N2 represent datasets with non-linearity.}
    \label{fig: first datasets}
\end{figure}

\begin{table}[t]
\small
\caption{the performance on linearly-separable datasets (L1 and L2) and non-linearly-separable datasets (N1 and N2)}
\label{tab:4 datasets}
\tabcolsep 9.5 pt
\begin{tabular}{|ccccc|}
\hline
\multicolumn{1}{|c|}{\multirow{2}{*}{Dataset}} & \multicolumn{2}{c|}{Classical}           & \multicolumn{2}{c|}{Quantum} \\
\multicolumn{1}{|c|}{}                         & Linear   & \multicolumn{1}{c|}{MLP}      & VQC-1           & VQC-2        \\ \hline
\multicolumn{1}{|c|}{L1}                       & 100\%    & \multicolumn{1}{c|}{100\%}    & 100\%         & 100\%        \\
\multicolumn{1}{|c|}{L2}                       & 99.688\% & \multicolumn{1}{c|}{100\%}    & 100\%         & 100\%        \\ \hline
\multicolumn{1}{|c|}{N1}                       & 70.625\% & \multicolumn{1}{c|}{100\%}    & 51.563\%      & 49.688\%     \\
\multicolumn{1}{|c|}{N2}                       & 66.250\% & \multicolumn{1}{c|}{96.875\%} & 47.188\%      & 49.375\%     \\ \hline
\end{tabular}%
\end{table}

For the development of quantum machine learning, it is necessary to build the correct scheme just like classical machine learning, while if we look at the history, the non-linearly activation function is a core part as it can help the network express and learn more complicated data. On the other hand, the quantum circuit does not provide such non-linear correlations, which prevents it from studying data and therefore building the non-linear mapping function while dealing with complex datasets. That is why it is so important to study non-linearity when trying to gain essential progress in the combination of quantum computing and machine learning. However, one challenge here is the need for proper datasets that could be used to examine the Non-linearity (i.e., able to perform well on datasets that are not linearly separable). In this work, we build valid datasets with proper size as shown in Figure \ref{fig: first datasets}. Experimental Setup in Section \MakeUppercase{\romannumeral 6} gives more details.


Table~\ref{tab:4 datasets} shows the performance of the different models, linear classifier, multilayer perceptron (MLP), and VQCs on the generated datasets. We observe that all models can get high performance on the linearly separable dataset, however, only MLP is capable to work on the non-linearly separable datasets. 
The key problem we need to figure out is why MLP can achieve the highest performance on non-linear datasets. One main reason should be the activation functions used in its hidden layers are non-linear. Common activation functions such as the Sigmoid, Tanh, and ReLU are all non-linear. The use of non-linear activation functions enables MLP to model complex non-linear relationships between inputs and outputs, leading to better performance on many tasks compared to linear models.  
 After figuring out the core reason why MLP can achieve non-linearly in a classification task, we need to actually feed non-linearity into the design of VQC to achieve similar performance. 

\subsection{Low Fidelity in Data Encoding Methods}
Amplitude encoding refers to normalizing the input data and then encoding them into the system's quantum state, which means that for $n$ input data, amplitude encoding needs $\log n$ qubits. While angle encoding, on the other hand, utilizes parameterized gates such as RX gate to encode the input data, thus $n$ qubits are required to encode n input data. However, there exists the same scalability issue in both methods while actually running in real quantum computers. Although amplitude encoding can express $2^n$ data with $n$ qubits, it has to be decomposed into more basic gates when deploying in the real quantum device, therefore resulting in more errors as the encoding circuit's length increases. For random data from MNIST, we tried different ways of amplitude encoding, ranging from 4-dimension data encoded in 2 qubits to 64-dimension data encoded in 6 qubits, and tested them on real quantum devices to see how many errors can be raised caused during the encoding circuit. As shown in Table~\ref{tab:Table 2}, as the number of encoding qubits increases, the circuit length starts growing which raises the average errors due to noise. 

\begin{table}[t]
\small
\caption{comparison of average errors as amplitude encoding qubits increases}
\label{tab:Table 2}
\tabcolsep 5.5 pt
\begin{tabular}{|c|cc|cc|}
\hline
\multirow{2}{*}{\# Qubits} & \multicolumn{2}{c|}{ibm\_geneva (27 qubits)}                    & \multicolumn{2}{c|}{ibmq\_jakarta (7 qubits)}                  \\ \cline{2-5} 
                                    & \multicolumn{1}{c|}{Circ. Len.} & Average Err. & \multicolumn{1}{c|}{Circ. Len.} & Average Err. \\ \hline
2                                   & \multicolumn{1}{c|}{11}             & 2.3450        & \multicolumn{1}{c|}{11}             & 0.7143        \\ \hline
3                                   & \multicolumn{1}{c|}{36}             & 1.9496        & \multicolumn{1}{c|}{35}             & 2.6842        \\ \hline
4                                   & \multicolumn{1}{c|}{102}            & 6.6860        & \multicolumn{1}{c|}{91}             & 8.1201        \\ \hline
5                                   & \multicolumn{1}{c|}{220}            & 10.8722       & \multicolumn{1}{c|}{211}            & 13.3838       \\ \hline
6                                   & \multicolumn{1}{c|}{478}            & 11.9897       & \multicolumn{1}{c|}{495}            & 20.4578       \\ \hline
\end{tabular}%

\end{table}

\begin{table}[t]
\small

\caption{the comparison between amplitude encoding and angle encoding on mnist dataset with 8$\times$8 dimension using 6 qubits }
\label{tab:Table 3}
\tabcolsep 15.5 pt
\renewcommand{\arraystretch}{1.1}
\begin{tabular}{|c|c|c|}
\hline
        & Amplitude encoding & Angle encoding \\ \hline
2-class & 97.5\%             & 90.0\%         \\ \hline
3-class & 92.5\%             & 73.0\%         \\ \hline
\end{tabular}%
\end{table}

Speaking of angle encoding, although it does not require decomposition in real quantum computers, it can only represent $n$ data with $n$ qubits compared to amplitude encoding. That leads to the situation where we normally have to use multiple gates in one qubit wire to encode multiple data, which would cause information loss. To prove it, we use 8$\times$8 input data from MNIST to do both a two-classification task and a three-classification task by using amplitude encoding and angel encoding separately. Since the dimension is 8$\times$8, amplitude encoding requires exactly 6 qubits, while angle encoding would need 64 qubits and this is not accepted. In this case, if we want to use less or the same number of qubits as amplitude encoding, we need to use multiple gates on one qubit. While based on the test result shown in Table~\ref{tab:Table 3}, angle encoding can only achieve 90.0\% accuracy while amplitude encoding can reach up to 97.5\%., and the difference in the accuracy due to information loss should be more intuitive on the three-classification task. Just to notice here, the experiment is held on a simulator instead of an actual quantum device.

Amplitude encoding has low fidelity while deploying in real quantum devices; on the other hand, Angle encoding would cause information loss when trying to maintain scalability. Motivated by that, we need to find a method that can make a good trade-off between scalability and fidelity to best protect the data from error and information loss to achieve better performance. 

\subsection{Lack of Spatial Features Extraction from the Existing Quantum Neural Network Designs}
\begin{figure}[t]
    \centering
    \includegraphics[width=1\columnwidth]{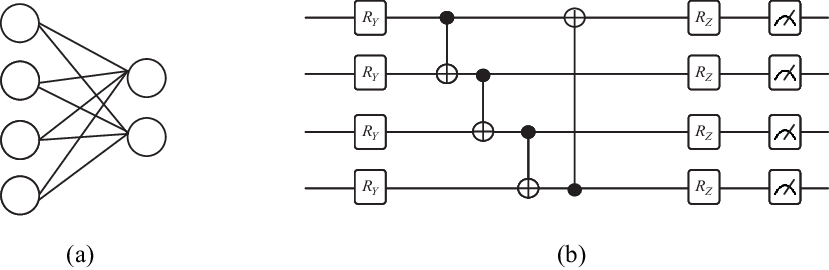}
    \caption{(a) A fully-connected linear classical neural network (b) A linear variational quantum neural network}
    \label{fig: linearty}
\end{figure}

If we take a look at the difference between the traditional fully connected layer and the convolution layer, we should notice the reason why CNN can easily handle image classification tasks is its ability to extract the spatial information from the original image. While the existing designs of VQC do not consider the locality but combine all information together. As shown in Figure~\ref{fig: linearty}, we can see that the existing VQC circuit is similar to a full-connected layer given that it involves all qubits in the computation at the same time, while on the other hand, the pattern of designing a quantum circuit that follows the pattern of classical image processing still lacks.

Motivated by that, we feel it necessary to introduce a new pattern of designing computation circuits, which can involve the concept of 'layer' to extract the features from input data gradually instead of computing the encoded data as a whole and ignoring the spatial information.

\subsection{Topology-Agnostic Design Brings Extra SWAP Gates to be Vulnerable to Quantum Noise}



Meanwhile, errors are raised due to the noise when we actually implement the logic quantum computation circuit on a real quantum device. 
The case here is that the execution of the circuit on the simulator looks great, however, after actually deploying the parameterized circuit on the real quantum computer, there is a heavy reduction in the performance due to the physic qubits’ noise. As shown in Figure \ref{fig: search result}, the accuracy of both amplitude and angle encoding drops less than 50.0\%. 

One main reason is that a physical qubit can only interact with its adjacent qubits. If there is an interaction such as a CRX gate between two logic qubits while their corresponding physic qubits are not neighbors, extra SWAP operation is required to swap the information which results in more CNOT gates and therefore brings more connection errors. 
 Thus, we can say that the direct mapping method in the existing design is vulnerable to quantum noise due to the extra SWAP gates in the computation circuit when running in a real quantum device.
It is important to also consider the actual topology of a specific real quantum device while designing the logic circuit to avoid unnecessary connection errors.

All of those challenges and motivations above inspire us to design and build a new Variational Quantum Circuit system to actually enable quantum deep learning and have better performance on real quantum processors.
\section{design}

\subsection{ST-VQC and Its Design Framework}
\begin{figure*}[t]
\centering
\includegraphics[width=1\linewidth]{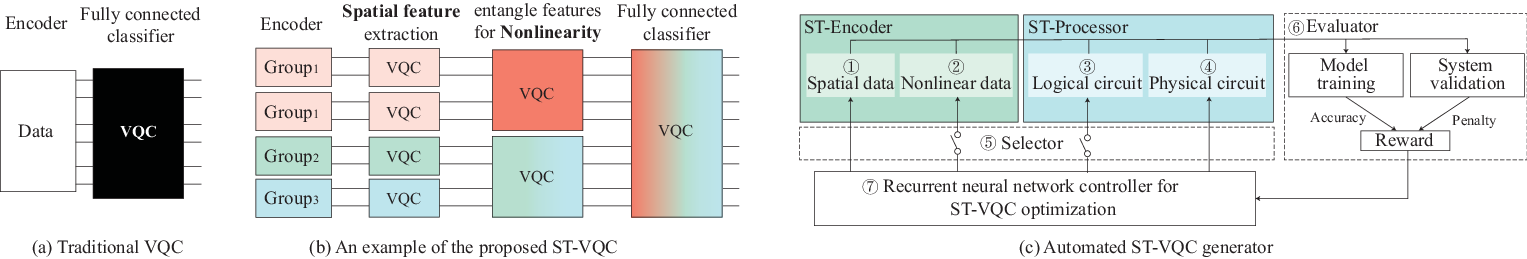}
\caption{From quantum ``shallow'' learning to ``deep'' learning: (a) the conventional variational quantum circuit (VQC); (b) the proposed ST-VQC with non-linearity to enable deep learning; (c) two proposed design blocks in the holistic optimization loop.}
\label{fig:search framework}
\end{figure*}

In order to address the challenges mentioned in Section \MakeUppercase{\romannumeral 3}, we propose a novel Spatial-Temporal Variational Quantum Circuit design, namely \textbf{ST-VQC}, as shown in Figure \ref{fig:search framework}(b).
Unlike the traditional VQC (Figure \ref{fig:search framework}(a)) to perform shallow learning, ST-VQC can extract the spatial feature and equip the ability to achieve non-linearity.
A holistic framework to generate an ST-VQC is shown in Figure \ref{fig:search framework}(c).

There are two core design blocks in the framework: \textbf{ST-Encoder} and \textbf{ST-Processor}.
ST-Encoder provides the data encoder circuit design to encode classical data to quantum computers; while ST-Processor provides the computation quantum circuit design.
There are two components in each design block.
In ST-Encoder, we propose a {\large\ding{192}} \textit{Spatial Data Encoder}, which has the ability to maintain the integrity of input data, reduce the needed number of qubits, and improve the fidelity.
The second component, {\large\ding{193}} \textit{Nonlinear Data Encoder}, is designed to integrate non-linearity into the forward propagation process of quantum neural networks. 

ST-Processor is designed to leverage the proposed data encoder, ST-Encoder, to extract spatial features and integrate non-linearity in computing.
Specifically, in {\large\ding{194}} \textit{Logical Circuit Design}, it can extract local spatial features while processing data with a cascaded structure in temporal. 
Then, the {\large\ding{195}} \textit{Physical Circuit Design} aims to avoid using the SWAP gate in the data processing phase, which may be involved due to no communication channel between two physical qubits and the CNOT gate is placed on these two qubits in logical circuit design.

The design of the ST-Encoder and ST-Processor will involve hyperparameters, which will affect the usage of resources.
With the goal of providing an automated design for quantum learning users, we proposed a holistic optimization framework, which will further include {\large\ding{196}} optimization path selector, {\large\ding{197}} a recurrent neural network based optimization controller, and {\large\ding{198}} a solution evaluator.
In this section, we will present the details of the ST-Encoder and ST-Processor. On top of understandings of the ST-VQC design, we will then give the automated system optimization in the next section.

\subsection{ST-Encoder: A Novel Data Encoder}

\vspace{1pt}
\noindent\textbf{{\large\ding{192}} Spatial Data Encoder}
\vspace{1pt}

\begin{figure}[t]
    \centering
    \includegraphics[width=1\columnwidth]{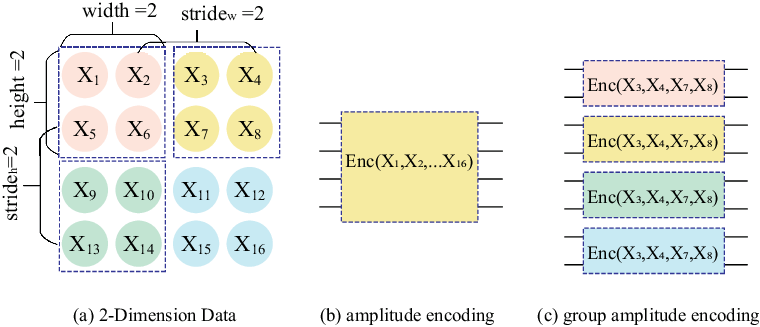}
    \caption{Illustration of spatial data encoder in ST-Encoder}
    \label{fig: data_group}
\end{figure}

In Section \MakeUppercase{\romannumeral 3}, we have demonstrated that both angle encoding and amplitude encoding cannot achieve high performance on real quantum devices. Since the angle encoding will cause information loss while encoding the mass of data on a limited number of qubits, and the usage for amplitude encoding will insert a large number of CNOT gates, we'd like to make a trade-off between these two encoding methods, which is called Spatial Data Encoder. 


\textbf{Design:}
The proposed encoder will collect the location-close data to groups, and map each group with $N$ data to $\lceil log_2{N}\rceil$ qubits using amplitude encoding.
For the example in Figure \ref{fig: data_group}, we are going to encode a $4 \times 4$ data.
If we apply amplitude encoding, it requires 4 qubits to map all data to the amplitudes, say the first data will be mapped to the amplitude of $|0000\rangle$.
On the other hand, if we apply angle encoding, to avoid information loss, it requires 16 qubits.
Now, let's see the proposed Spatial Data Encoder in  Figure \ref{fig: data_group}(c).
It creates 4 groups with $2 \times 2$ data, where each group with 4 data can be encoded into 2 qubits, and thus it needs 8 qubits total.

\textbf{Design Space:}
In Spatial Data Encoder, it has the flexibility to change the shape of groups to adjust the required CNOT gates, and also, we allow to have overlap between groups.
Kindly note that compare with amplitude encoding and angle encoding, the proposed Spatial Data Encoder is more general.
When the number of groups is 1, it is equivalent to amplitude encoding; while, if the number of elements in each group is 1, it is equivalent to angle encoding.
Formally, we define a group with 3 hyperparameters, denoted as $f=(W,H,S)$, where $W$ is the width, $H$ is the height of the group data (or called ``filter'') and $S$ is the step size of the filter. After the parameters $f =(W,H,S)$ are determined, the number of groups $g$ will be fixed.

\vspace{1pt}
\noindent\textbf{{\large\ding{193}} Nonlinear Data Encoder}
\vspace{1pt}

\begin{figure}[t]
    \centering
    \includegraphics[width=1\linewidth]{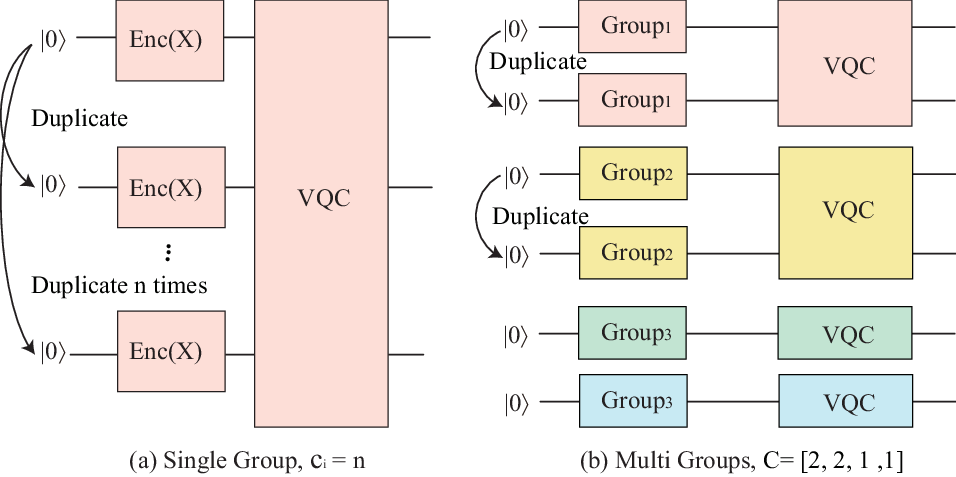}
    \caption{Illustration of nonlinear data encoder in ST-Encoder}
    \label{fig: Data Duplicate}
\end{figure}

The tensor product ($\otimes$) in quantum circuits refers to the process of composing two or more quantum systems into a single, larger system. It is a fundamental operation in quantum computing and is used to model the interactions between multiple qubits or to perform complex quantum algorithms. The resulting system has the same number of qubits as the original systems combined, and the state of each qubit in the new system is a tensor product of the states of the corresponding qubits in the original systems.
For example,
Suppose we have two single-qubit states:
\begin{equation}
\begin{split}
 |\psi_1\rangle = \alpha|0\rangle + \beta|1\rangle \\
 |\psi_2\rangle = \gamma|0\rangle + \delta|1\rangle 
\end{split}
\end{equation}
To form the tensor product of these two states, we simply combine them by placing one state after the other:
\begin{equation}
\begin{split}
|\psi_1\rangle \otimes |\psi_2\rangle 
&= (\alpha|0\rangle + \beta|1\rangle) \otimes (\gamma|0\rangle + \delta|1\rangle) \\
&= \alpha\gamma|00\rangle + \alpha\delta|01\rangle + \beta\gamma|10\rangle + \beta\delta|11\rangle
\end{split}
\label{eq:tensorproduct}
\end{equation}

\textbf{Design:}
Since the data are encoded on the amplitudes, our nonlinear data encoder is designed based on the above property.
More specifically, to achieve a quadratic non-linearity, we only need to make $\alpha=\gamma$ and $\beta=\delta$ in the above equations, which is to duplicate one same input data to two qubits.
Let $\alpha = \gamma = x_1 $ and $\beta = \delta = x_2 $, the Equation~\ref{eq:tensorproduct} becomes:
\begin{equation}
\begin{split}
    |\psi_1\rangle \otimes |\psi_2\rangle 
    &= (x_1|0\rangle + x_2|1\rangle) \otimes (x_1|0\rangle + x_2|1\rangle) \\
    &= x_1^2|00\rangle + x_1x_2|01\rangle + x_1x_2|10\rangle + x_2^2|11\rangle
\end{split}
\end{equation}
In the above equation, it is clear that the quadratic terms $x_1^2$, $x_1x_2$, $x_2^2$ are introduced in the quantum system.

By duplicating input data $n$ times, we can include the n-order non-linearity in the quantum learning circuit, as shown in  Figure \ref{fig: Data Duplicate}(a).
Kindly note that there will be a trade-off between the non-linearity order and the number of qubits.

\textbf{Design Space:} Given a quantum circuit with spatial data encoder (see Figure \ref{fig: data_group}), we use a vector of hyperparameters $C = \{c_1,c_2,\cdots,c_g\}$ to indicate the number of duplications for each group, where $g$ is the number of group structure and each element  $c_i$ is the number of duplications for the $i^{th}$ group.
As an example, in Figure \ref{fig: Data Duplicate}, we show the duplicated circuit with parameter $C=\{2,2,1,1\}$ of Figure \ref{fig: data_group}.
Kindly note that the range of hyperparameters is limited by the total number of physical qubits.
Let $q\_i\in \{q_1,q_2,\cdots,q_g\}$ denote the number of qubits used in the $i^th$ group, and $Q$ be the total number of qubits.
We have the constraint $\sum_{\forall i}\{c_i\times q_i\}\le Q$.



\begin{figure}[t]
    \centering
    \includegraphics[width=1\columnwidth]{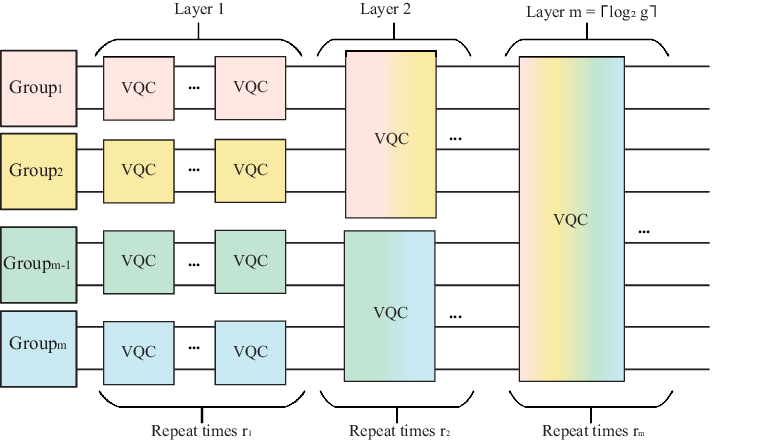}
    \caption{Local spatial and cascaded temporal logical circuit design}
    \label{fig: Layer-wise}
\end{figure}

\subsection{ST-Processor: Spatial-Temporal Data Processor}


\vspace{1pt}
\noindent\textbf{{\large\ding{194}} Local Spatial \& Cascaded Temporal Logical Circuit}
\vspace{1pt}

As discussed in Section \MakeUppercase{\romannumeral 3} \textit{C}, the conventional VQC design lacks the capability to extract spatial information, which is crucial in handling datasets with spatial information, such as image datasets. 
Likewise the development of convolution operation from fully connected operation in classical computing, we propose to implement a set of VQCs in a cascaded structure on time, which will include multiple layers. 

\textbf{Design:} For the data encoded by the nonlinear data encoder, we will first extract the local spatial information with a VQC on each group. Then, we will gradually enlarge the VQC to involve more qubits.
Along with the merge of groups, the non-linearity will be created, and we regard a new layer starts.
Finally, we will have a VQC to cover all qubits.
Such a design will form a reverse tree structure, as shown in Figure \ref{fig: Layer-wise}.
From this figure, we can see that after we split the data into groups, we will have a 3-layer VQC design. At layer 1, there are 4 small VQCs where each is responsible for one data group, therefore at this moment, only qubits within the same block will be entangled together and the information for this part of values will be extracted by the end of this layer. If we recall the CNN, we could consider this as one convolution layer that extracts spatial information from an image. Then, at layer 2, we have two VQCs and now each is responsible for two data blocks where 4 qubits are allowed to be entangled. Finally, at layer 3, we make sure of the entanglement among all qubits, and we have successfully built a framework similar to a deep learning network and extracted the spatial features from the input.

\textbf{Design Space:} In the above design, there exist a set of hyperparameters to control the length of the circuit.
Let $R = [r_1,r_2,...,r_m]$, where $m$ is the maximum number of layers and $r_i$ represents the repeated times for the  $i^{th}$ layer. 
The value of $m$ can be determined by the number of groups, that is, $m = \lceil log_2 g \rceil $. 
Kindly note that the output quantum circuit can have less than $m$ layers, when $\exists r_i\in R, s.t., r_i=0$.
A special case is that if $\forall r_i\in R$ and $i\ne m$, $r_i=0$, the outcome circuit is equivalent to the conventional VQC.


\noindent\textbf{{\large\ding{195}} SWAP-Free Physical Circuit Design}

\begin{figure}[t]
    \centering
    \includegraphics[width=1\columnwidth]{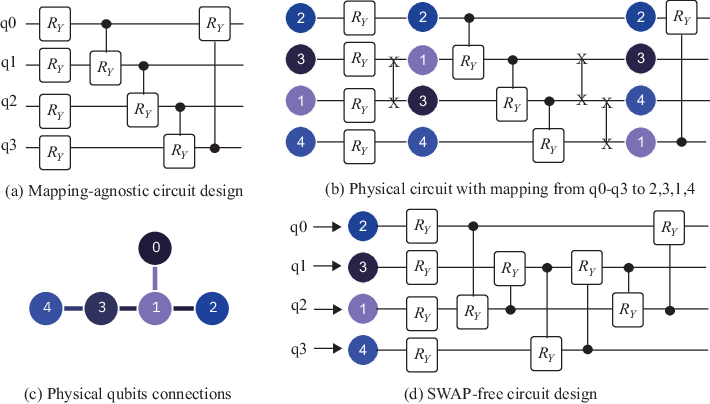}
    \caption{Illustration of advantage of SWAP-free circuit design: (a-b) mapping-agnostic circuit design will involve additional SWAP gates; (c) the topology of physical qubits; (d) the proposed SWAP-Free design.}
    \label{fig: mapping}
\end{figure}

We have demonstrated in Section \MakeUppercase{\romannumeral 3} \textit{D} that a VQC design with too many gates will lead to significant performance degradation due to noise.
However, the traditional variational quantum circuit shown in Figure~\ref{fig: mapping}(a) did not consider the topology of a real quantum device and the mapping from logical qubits to physical qubits.
As a result, as shown in  Figure~\ref{fig: mapping}(b), for a run-time determined mapping, the circuit needs a SWAP gate since the CNOT gate between a pair of logical qubits may not have the connection between the physical qubits to which they mapped.
As we know, each SWAP gate is further implemented by 3 CNOT gates, and therefore, additional CNOT gates will be involved in such a mapping-agnostic circuit design.

\textbf{Design:} We propose a SWAP-gate free design shown in Figure~\ref{fig: mapping}(d).
The basic idea is to design the VQC based on the mapping and the connectivity of different qubits.
Specifically, compared with the original VQC design in Figure~\ref{fig: mapping}(a), we made two-step modifications: (1) for the first three CNOT gates, we keep the property that the gates form a path but ensure all CNOT gates on the physical qubits have direct connections. As an example, we change it from $q_0\rightarrow q_1\rightarrow q_2\rightarrow q_3$ to $q_0\rightarrow q_2\rightarrow q_1\rightarrow q_3$.
Note that the circuit is changed before the training phase so that the logical function will not be changed for this modification step.
(2) for the last CNOT gate $q_3\rightarrow q_0$ in Figure~\ref{fig: mapping}(a), it aims to make entanglement between $q_3$ and $q_0$. However, it forms a circle with an exact number of nodes. In most cases, we cannot perfectly match such a circuit to a quantum device, and therefore SWAP will be involved.
To avoid the involvement of additional SWAP, we propose to entangle the tailor and the head of the previous path in an indirect way: we will go back from the tail along with the path toward the head.
In this case, if we can guarantee that the path in the first step has no SWAP gate, then, the full circuit does not need SWAP gates.

If the number of logical qubits approaches the physical qubits in a quantum processor, it is possible that no paths have a certain number of qubits.
In this case, we will notify the user to apply a quantum processor with more qubits. Or, we will help the user to minimize the SWAP gate.
The optimization will be carried out in four steps: (1) we will find a path $P$ including the maximum number of qubits; (2) we create a subgraph $G$ and initialize $G=P$; (3) we will iteratively add a qubit to $G$ such that the maximum length of the newly added qubits to the path $P$ is minimized, we denote these metrics as the ``fat'' of $G$; (4) after we obtain the sub-graph $G$, then, we will use a modified depth-first search algorithm \cite{tarjan1972depth} to place CNOT gates and SWAP on the logical circuit. The basic idea is that when we reach a qubit having a degree larger than 2, we will leave the path $P$. The search will keep going until we meet a leaf node or it reaches a node in $P$.

\textbf{Design Space:} At offline, for a given VQC with maximum $n$ qubits, we will use a depth-first search (DFS) algorithm\cite{tarjan1972depth} to create paths with less than or equal to $n$ qubits or a set of the subgraph that has the minimum ``fat'' metric, as candidates.
We will then calculate the sum of noise for each candidate, including all the single-qubit errors and CNOT errors. To reduce the search space, we build up a set $S$ that includes the top-k candidates for each possible number of qubits with lower noise levels than others.




\section{system-level optimization}\label{subsec:5.2}\label{subsec:5.1}

\textbf{Overview:} As the optimization loop shown in Figure \ref{fig: data_group}(c). 
In addition to the search spaces of ST-Encoder and ST-Processor introduced in the previous section, it further contains {\large\ding{196}} optimization path selector, {\large\ding{197}} a recurrent neural network based optimization controller, and {\large\ding{198}} a solution evaluator.
In each episode (a.k.a. iteration), the optimizer selector first determines the search space, then the controller samples the solution from the search space, and the predicted sample goes through the evaluator to generate the accuracy and penalty. Finally, a reward is generated to update the controller. All the components work together to generate solutions with high-weighted accuracy and meet all design constraints. To realize the ST-VQC, we apply the reinforcement learning approach in this paper. Based on the formulated reward function, other optimization approaches, such as evolution algorithms, can also be applied. 
In the following text, we will introduce the four search spaces relative to four designs in detail.

\vspace{1pt}
\noindent\textbf{{\large\ding{196}} Optimizer Selector}
\vspace{1pt}

We integrate an optimizer selector in
ST-VQC to make it adapt to different kinds of datasets. If the dataset is a linear-separable dataset, such as MNIST, we will turn off the component of `Non-linearity Exploration'. Since the layer-wise structure is designed for extracting spatial information, we can turn off the component of `Layer Structure Exploration' when we process vector-like data, such as text. The absence of components will reduce the search space, and therefore, the exploration cost will also be reduced.

\vspace{1pt}
\noindent\textbf{{\large\ding{197}} Evaluator}
\vspace{1pt}

The evaluator is a key component in the framework of quantum neural network design.
The evaluator contains four components: training, mapping, evaluation, and cost model. In each episode, the co-exploration controller predicts the design of the QNN model and mapping strategy, and the QNN model will be processed in the evaluator to build up reward and penalty. 
The evaluator contains two paths: (1) via the training, mapping, and validating to obtain VQC accuracy on the real quantum device; (2) via cost modeling to generate penalty in terms of design specs and qubit resource limitation. 

\textit{Model training.} 
Hyperparameters for QNN's design, including {\large\ding{192}} group structure, {\large\ding{193}} duplication times for nonlinearity and {\large\ding{194}} layer structure are obtained from the controller. The QNN model is trained on a classical computer first, then the logical design will be mapped to physical qubits according to the predicted {\large\ding{195}} mapping.
Note that our proposed ST-VQC is to automatically build up a design of QNN on the real quantum device to get high performance, and therefore the final validation is on the real quantum device. The accuracy on the validation dataset is used for the reward $R$.

\begin{figure}[t]
    \centering
    \includegraphics[width=1\columnwidth]{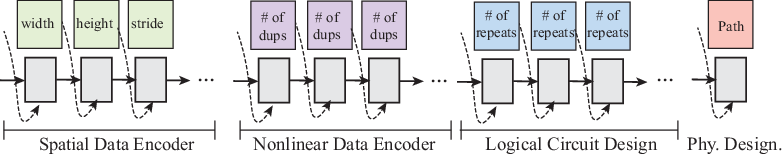}
    \caption{Exploring search space using an RNN controller}
    \label{fig: search space}
\end{figure}

\textit{System validation.}
After we have a solution, we need to validate whether it can be deployed to the given quantum devices with limited qubits resources.
If it is invalid, we will add a penalty to the reward function.
To construct the penalty, we will create two binary variables.
The first constraint $n \leq N $ represents the limitation of qubit resource, where $n$ is the number of qubits needed by the design and $N$ is the number of qubits on the longest path of the given quantum device $D$. 
We create a binary indicator $bn=0$, if $n\leq N$, otherwise $bn=1$.
The second constraint is the number of qubits in the identified subgraph $q\in S$ should be the same as the number of qubits $n$ in the identified VQC.
We create another binary indicator $bq=0$, if $|q|\ne n$, otherwise $bq=1$.
Then, the penalty can be formulated as follows.
\begin{equation}
    P = bn + bq
\end{equation}

\textit{Reward function.} 
Based on all the above evaluation results, we calculate the reward with a scaling variable $\rho$, listed as follows:
\begin{equation}
    R(D, P) = acc - \rho \times P
\end{equation}

\noindent\textbf{{\large\ding{198}} Reinforcement Learning Controller}

We propose to use a reinforcement learning-based Recurrent Neural Network (RNN) controller to sample the solution from the search space. 
Figure~\ref{fig: search space} demonstrates the proposed controller. 
It is composed of 4 segments, including spatial data encoder, nonlinear data encoder, logical circuit design, and physical design. Each segment predicts the corresponding solution.
For instance, in Figure~\ref{fig: search space}, the first segment predicts the hyperparameter of the spatial data encoder, including the height, weight, and strides of the group data (or called "filter").

We employ the reinforcement learning method to update the controller and predict new samples. Specifically, in each episode, the controller first predicts a sample and gets its reward $R$ based on the evaluation results. Then, we employ the Monte Carlo policy gradient algorithm \cite{williams1992simple} to update the controller: 
\begin{equation}
\nabla J(\theta)=\frac{1}{m} \sum_{k=1}^m \sum_{t=1}^T \gamma^{T-t} \nabla_\theta \log \pi_\theta\left(a_t \mid a_{(t-1): 1}\right)\left(R_k-b\right)
\end{equation}
where $m$ is the batch size and $T$ is the number of steps in each episode. Rewards are discounted at every step by an exponential factor $\gamma$ and the baseline $b$ is the exponential moving average of rewards.

\section{experimental results}
\begin{table}[t]
\centering
\caption{evaluation of different quantum designs for mnist-2 on \textit{ibm\_cairo} quantum processor}
\label{tab:exp1}
\small
\tabcolsep 5 pt
\renewcommand{\arraystretch}{1.1}{%
\begin{tabular}{|c|c|c|c|c|c|}
\hline
ID & Method        & \#Qubits & Cir. Depth & \#Para. & Accuracy \\ \hline
S1           & Amplitude VQC & 4        & 422       & 40      & 47.5\%   \\ \hline
S2           & Angle-4 VQC     & 4        & 226       & 40      & 45.0\%   \\ \hline
S3           & Angle-8 VQC     & 8        & 233       & 44      & 52.5\%   \\ \hline
S4           & ST-VQC (Acc)        & 8        & 309       & 54      & 90.0\%   \\ \hline
S5           & ST-VQC (Qub)        & 6        &    191    & 34      & 80.5\%   \\ \hline
\end{tabular}%
}
\end{table}

\begin{table}[t]
\caption{evaluation of different encoding methods for mnist-2 on \textit{ibm\_cairo} quantum processor}
\label{tab:exp2}
\small
\tabcolsep 3.5 pt
\renewcommand{\arraystretch}{1.1}{%
\begin{tabular}{|c|cc|c|c|c|}
\hline
ID & \multicolumn{1}{c|}{Encoding}  & Computation  & Enc. Depth & Deviation  & Accuracy \\ \hline
S6           & \multicolumn{1}{c|}{Amplitude} & ST-Processor & 99        & 0.0938 & 48.0\%   \\ \hline
S7           & \multicolumn{1}{c|}{Angle-4}     & ST-Processor  & 5         & 0.0032 & 72.5\%   \\ \hline
S4           & \multicolumn{1}{c|}{ST-Encoder}  & ST-Processor                 & 11        & 0.0058 & 90.0\%   \\ \hline
\end{tabular}%
}
\end{table}

\begin{table}[]
\centering
\caption{evaluation of different computation circuits for mnist-2 on \textit{ibm\_cairo} quantum processor}
\label{tab:exp3}
\small
\tabcolsep 8.5 pt
\renewcommand{\arraystretch}{1.1}
\begin{tabular}{|c|cc|c|c|}
\hline
ID & \multicolumn{1}{c|}{Encoding}   & Computation & \#CNOT & Accuracy \\ \hline
S8           & \multicolumn{1}{c|}{ST-Encoder} & VQC         & 147    & 58.0\%   \\ \hline
S4           & \multicolumn{1}{c|}{ST-Encoder} & ST-Processor                  & 94     & 90.0\%   \\ \hline
\end{tabular}%

\end{table}

\begin{figure}[t]
    \centering
    \includegraphics[width=1\linewidth]{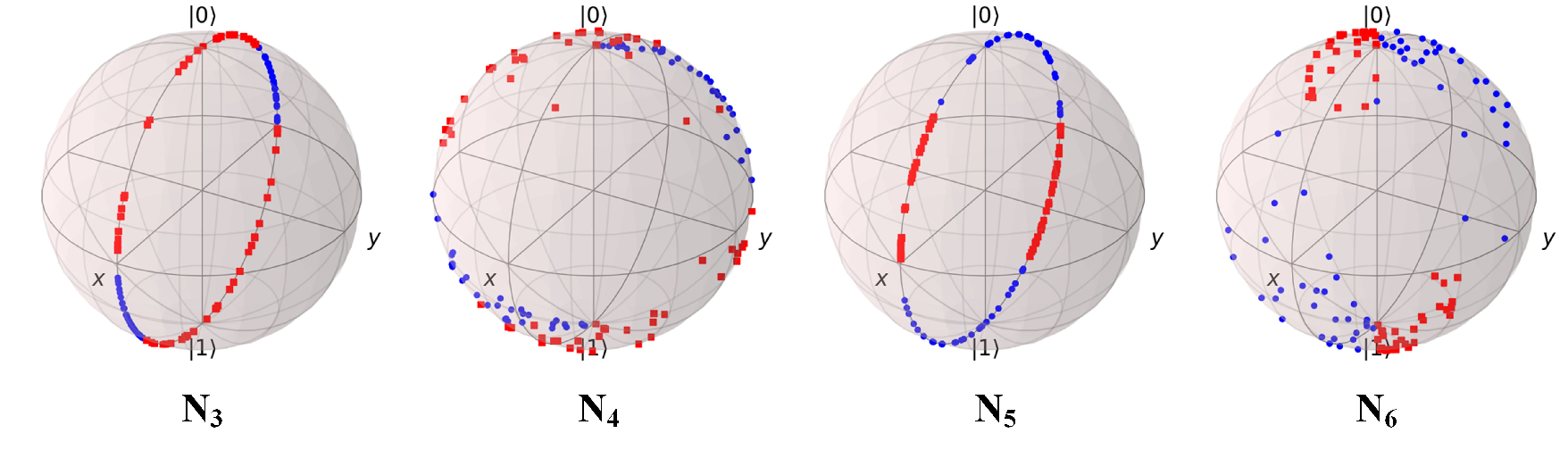}
    \caption{Four datasets(N3, N4, N5, N6) that are non-linearly separable used to evaluate existing quantum circuit models.}
    \label{fig: more datasets}
\end{figure}
\begin{figure}[t]
    \centering
    \includegraphics[width=1\columnwidth]{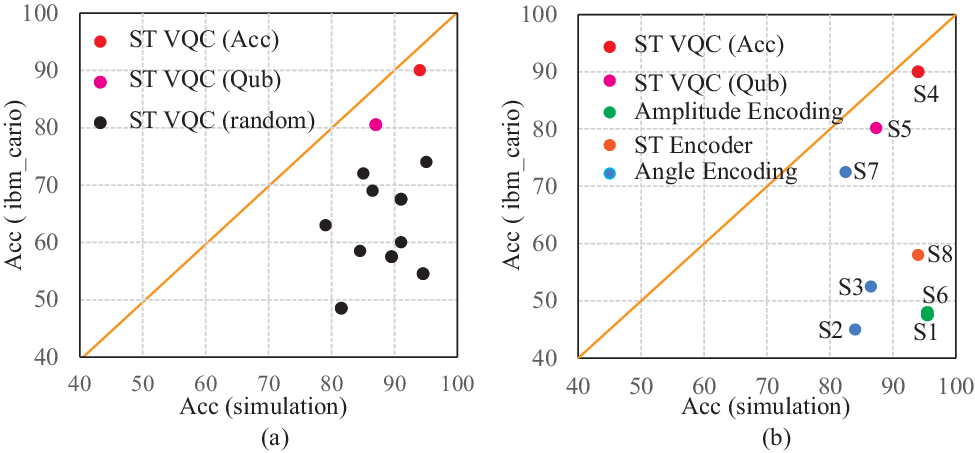}
    \caption{RL-baased optimization can outperform random solutions, and ST-VQC are more robust to noise against the existing VQCs. }
    \label{fig: search result}
\end{figure}

\subsection{Experimental Setup}

\textbf{Dataset.}
Our experiments are conducted on two datasets. Firstly we evaluate our proposed design on MNIST 2-Class(3, 6). The input images used in the experiments are downsampled to a dimension of 4$\times$4 from the original 28$\times$28. We only choose the first 200 images as the test samples due to the limited quantum resources.  

The presence of non-linearity can be shown in widely used datasets. Nevertheless, the capability of quantum machine learning poses certain limitations on the exploitation of some complex datasets. In order to better evaluate the effectiveness of the proposed ST-Encoder on processing non-linearity data, we generate a set of synthetic datasets followed by the ideas: while thinking of the non-linearity, the most straightforward example would be the XOR gate. It is simple to construct the classical XOR training dataset on the basis of input of two values with one output following the design of the XOR gate. However, in our case, we need to consider creating the dataset based on the quantum rotation. To that end, we build the dataset on the Bloch Sphere instead of using the Cartesian Coordinate Plane as shown in Figures \ref{fig: first datasets} and \ref{fig: more datasets}. To be more detailed, we use the parameterized rotation gates RY gate and U3 gate to adjust the quantum state of one qubit. The points will then be constrained to the specific areas on the surface of the Bloch Sphere by limiting the values of rotation gates $\theta$. After that, we control the distribution of the points and labels (i.e., classes) to create linear separable and non-linear separable datasets. To better evaluate our methods, we generate datasets N3 to N6 in Figure \ref{fig: more datasets} to have different distributions, compared with N1 and N2 in Figure \ref{fig: first datasets}. Here, we can see that points shown in Datasets L1, N1, N3, and N5 are all around on the RY plane. While if we treat the sphere as 8 blocks from the X, Y, and Z axis, Datasets L2, N2, N4, and N6 have points on the surface of each block. It is clear to see from the figure that L1 is linearly separable while N1 is not. For better visualization, we don't plot all points on L2, N2, N4, and N6. For L2 which is linearly separable, the red points and blue points take half of the sphere respectively. Then for non-linearly separable N2, each pair of blocks on the diagonal is in different classes.


\textbf{Details of Training.} All experiments are implemented and trained by using the Python Library TorchQuantum. Experiments on MNIST-2 are trained with batch size 128 for 20 epochs, using the Adam optimizer with an initial learning rate of 0.005.
Then, experiments on synthetic datasets are trained with batch size 128 for 50 epochs, using the same optimizer and initial learning rate.

\textbf{Details of Testing.} To best observe the models' performance on real quantum computers, we send our MNIST-2 dataset test samples to the real quantum device ibm\_cario in IBM quantum with 27 qubits. Its quantum volume is 64 and the circuit layer operations per second are 2.4k. Then, we evaluate the Synthetic dataset test samples on a real quantum device ibmq\_lima with 5 qubits. Its quantum volume is 8 and the circuit layer operations per second are 2.7k.
Kindly note that it takes average 10 hours in queue for one model execution on ibm\_cario, and all experiments on MNIST-2 are performed on this quantum processor.

\textbf{Baselines.}
For MNIST-2 experiments, multiple models with different settings serve as the baseline. Notice that there are many possible ways of applying angle encoding. In this experiment, we use 4$\times$4\_ryzxy and 8$\times$2\_ryz encoding operator lists to do the angle encoding with 4 qubits and 8 qubits separately. 4$\times$4\_ryzxy refers to the repeated use of RY, RZ, and RX gates to encode every 4 values to 4 qubits until all 16 values are encoded. Similarly, 8$\times$2\_ryz uses repeatedly RY and RZ gates to encode every 8 values to 8 qubits until all 16 values are encoded.
For synthetic dataset experiments, we use two classical learning models: (1) 2$\times$2 Linear Classifier as the baseline, and (2) Multi-layer Perceptron (MLP) with one hider layer and uses the quadratic function as the activation function.

\subsection{Evaluation of ST-VQC on MNIST-2}

\textbf{Evalution of ST-VQC.} Table~\ref{tab:exp1} shows the evaluation of 5 different quantum designs for MNIST-2 on \textit{ibm\_cairo} quantum processor.
Angle-4 VQC and Angle-8 VQC refer to the design with angle encoding using 4 and 8 qubits, respectively.
Amplitude VQC refers to the design with amplitude encoding.
All these existing designs have the same and typically used backbone architecture, i.e., Circuit 18 in \cite{sim2019expressibility}.
We compare two solutions generated by the ST-VQC framework: (1) ST-VQC (Acc) with the highest accuracy, and (2) ST-VQC (Qub) with the limited number of qubits.

In Table~\ref{tab:exp1}, the column ``Cir. Depth'' corresponds to the total circuit depth for data encoder and computation, and the column ``\# Para.'' shows the number of parameters in the circuit.
From the table, it is clear that ST-VQC outperforms the other existing VQC on real quantum computers significantly. 
More specifically, the accuracy of the existing quantum designs can be regarded as a random guess, which are $45\%$, $47.5\%$, and $52.5\%$ for Angle-4 VQC, Amplitude VQC, and Angle-8 VQC, respectively. 
On the other hand, ST-VQC (Acc) and ST-VQC (Qub) can achieve an accuracy of $90\%$ and $80.5\%$, respectively.
An interesting observation is that ST-VQC (Qub) has the least number of parameters and shortest circuit depth; however, its accuracy is much higher than the existing VQC.
This demonstrates the importance of a system-level optimization to address the deployability issue for quantum learning algorithms.

To better investigate where the advantage of ST-VQC came from, we conduct two ablation studies as follows.

\textbf{Evaluation of ST-Encoder.} Table~\ref{tab:exp2} shows the evaluation of different encoding methods for MNIST-2 on \textit{ibm\_cairo} quantum processor, where we apply the same computation circuit proposed by ST-Processor for amplitude encoding, angle encoding with 4 qubits, and the proposed ST-Encoder. The column ``Enc. Depth'' corresponds to circuit depth for data encoding. And ``Deviation'' is a metric to calculate the difference between the output from the ideal simulator without noise and \textit{ibm\_cairo} quantum processor.


Table~\ref{tab:exp2} shows that ST-Encoder outperforms both angle and amplitude encoding. Specifically, compared to amplitude encoding, ST-Encoder has a shorter encoding circuit and therefore results in a $42\%$ increase in accuracy due to the smaller deviation in the output. Despite having a slightly longer encoding circuit that results in a larger deviation, ST-Encoder still provides an $18.5\%$ accuracy improvement over angle encoding. This is because angle encoding needs to encode multiple data to one qubit, which will cause information loss on features.
The above results illustrate the superiority of our proposed data encoding method.

 

\textbf{Evaluation of ST-Processor.} Table~\ref{tab:exp3} shows the evaluation of different computation circuits for MNIST-2 on \textit{ibm\_cairo} quantum processor, where ST-Encoder is used for data encoding in both cases. The computation circuit for S7 is manually designed while the computation circuit for S4 is proposed by ST-Processor. Note that the column ``\#CNOT'' is the number of CNOT gates after compilation, where the insertion of a single swap gate during compilation could introduce three extra CNOT gates. 

Results in Table~\ref{tab:exp3} show that the computation circuit designed by ST-Processor can improve the accuracy by a large margin on real quantum computers. More specifically, ST-Processor improves the accuracy by $32\%$ compared with the manually designed VQC since it has a much shorter encoding circuit. 
The same observation can be also got from the comparison between solutions S2 and S7. They have the same encoding method, Angle-4, but S7 applies our proposed ST-Processor, it can achieve a $27.5\%$ accuracy gain. 
SWAP-Free physical circuit design in the ST-Processor is evident in the large reduction in the number of CNOT gates.

\textbf{Evaluation of ST-VQC on ideal quantum simulator vs. real quantum computer.}

Figure~\ref{fig: search result} shows the evaluation of ST-VQC on the ideal quantum simulator (w/t noise) and \textit{ibm\_cairo} quantum processor (w/ noise), where the black points represent the circuit design which is randomly sampled from the search space defined in Section~\ref{subsec:5.1}.
Kindly note that we tried to apply the random approach to generate 50 solutions, but most of them are invalid because the numbers of their required qubits exceed the available qubits of \textit{ibm\_cairo}.
We also recorded the simulation results of other solutions in Tables \ref{tab:exp1}, \ref{tab:exp2}, and \ref{tab:exp3} for comparison.
In these figures, the red and pink points correspond to solutions S4 and S5 generated by the optimization in Section~\ref{subsec:5.2}. The solid orange line is a reference, which indicates that the simulation accuracy and accuracy on the real quantum devices are the same, which can hardly be achieved in practice due to noise. 

From Figure~\ref{fig: search result}(a), it is obvious that both red and pink points are closest to the solid orange line, which indicates that the accuracy of the corresponding circuit design on real quantum computers is closest to that on the ideal quantum simulator. Therefore, we can conclude that ST-VQC with system-level optimization can find the circuits that are robust to noise.

Comparison results with existing VQCs are reported in Figure~\ref{fig: search result}(b).
From this figure, we can see that although solutions S1, S6, and S8 can achieve high simulation accuracy, they suffer from a significant accuracy drop after deploying to the real quantum computers.
This reveals the deployability issue in the existing VQCs and again emphasizes the superiority of ST-VQC (S4 and S5) in providing high robustness.



\subsection{Evaluation on Nonlinear Datasets}

Table~\ref{tab:exp4} shows the evaluation of ST-VQC on nonlinear datasets N1-N6. The columns ``Linear'' and ``MLP'' refer to the linear classifier and multilayer perceptron running on classical computers, respectively. The performance of MLP serves as the upper bound of the performance on synthetic datasets. The column ``VQC'' and the column ``ST-VQC'' under ``simulator'' report the performance on the ideal simulator while the two columns under \textit{ibmq\_lima} report the performance executed on \textit{ibmq\_lima} quantum processor. 

In Table~\ref{tab:exp4}, it is obvious that ST-VQC can make accurate inferences on the nonlinear dataset since it introduces non-linearity through ST-Encoder, which cannot be handled by VQC. Specifically, on the ideal simulator, ST-VQC can achieve an accuracy of $94.71 \%$ on average while VQC can only have an accuracy of $49.53 \%$ on average. Moreover, the accuracy of ST-VQC on ibmq\_lima is close to that of the ideal simulator, which only incurs an accuracy loss of $1.7 \%$ on average. Besides, the accuracy of ST-VQC is comparable with that of MLP on classical computers. More specifically, the average accuracy loss is $2.32 \%$ and $4.02 \%$ on the ideal simulator and \textit{ibmq\_lima}, respectively.

\begin{table}[t]
\caption{evaluation on nonlinear datasets}
\label{tab:exp4}
\small
\tabcolsep 2.7 pt
\renewcommand{\arraystretch}{1.1}{%
\begin{tabular}{|c|ccclcclc|}
\hline
\multirow{2}{*}{Dataset} & \multicolumn{2}{c|}{Classical}           & \multicolumn{3}{c|}{Simulator}                                 & \multicolumn{3}{c|}{ibmq\_lima}               \\
                         & Linear   & \multicolumn{1}{c|}{MLP}      & \multicolumn{2}{c}{VQC}      & \multicolumn{1}{c|}{ST-VQC} & \multicolumn{2}{c}{VQC}     & ST-VQC      \\ \hline
N1                       & 70.63\%  & \multicolumn{1}{c|}{100.00\%} & \multicolumn{2}{c}{51.56\%}  & \multicolumn{1}{c|}{99.38\%}    & \multicolumn{2}{c}{49.38\%} & 98.13\%         \\
N2                       & 66.25\%  & \multicolumn{1}{c|}{96.88\%}  & \multicolumn{2}{c}{47.19\%}  & \multicolumn{1}{c|}{96.56\%}    & \multicolumn{2}{c}{45.69\%} & 95.50\%         \\
N3                       & 71.88\%  & \multicolumn{1}{c|}{98.44\%}  & \multicolumn{2}{c}{53.44\%}  & \multicolumn{1}{c|}{96.25\%}    & \multicolumn{2}{c}{53.75\%} & 93.69\%         \\
N4                       & 65.94\%  & \multicolumn{1}{c|}{96.25\%}  & \multicolumn{2}{c}{50.94\%}  & \multicolumn{1}{c|}{91.75\%}    & \multicolumn{2}{c}{48.38\%} & 90.00\%         \\
N5                       & 64.69\%  & \multicolumn{1}{c|}{93.75\%}  & \multicolumn{2}{c}{48.13\%}  & \multicolumn{1}{c|}{90.63\%}    & \multicolumn{2}{c}{46.69\%} & 89.50\%         \\
N6                       & 51.25\%  & \multicolumn{1}{c|}{96.88\%}  & \multicolumn{2}{c}{45.94\%}  & \multicolumn{1}{c|}{93.69\%}    & \multicolumn{2}{c}{44.13\%} & 91.75\%         \\ 
\hline
Avg.                       & 65.11\%  & \multicolumn{1}{c|}{97.03\%}  & \multicolumn{2}{c}{49.53\%}  & \multicolumn{1}{c|}{94.71\%}    & \multicolumn{2}{c}{48.00\%} & 93.01\%         \\ 
Imp.                       & baseline  & \multicolumn{1}{c|}{31.9\%}  & \multicolumn{2}{c}{-15.58\%}  & \multicolumn{1}{c|}{29.6\%}    & \multicolumn{2}{c}{-17.11\%} & 27.9\%         \\

\hline
\end{tabular}%
}
\end{table}
\section{conclusion}
There faces challenges in quantum learning: the lack of non-linearity and spatial feature extraction, as well as low fidelity in design and deployment. To address these, we provide a system-level automatic design framework. The framework is composed of two core design blocks: ST-Encoder and ST-Processor.
In ST-Encoder, we propose a nonlinear data encoder to integrate non-linearity into the forward propagation process of quantum neural networks and a spatial data encoder to make the trade-off between qubits number and fidelity. In ST-Processor, we propose local spatial \& cascaded temporal logical circuit design and a SWAP-free physical circuit design to extract spatial features and improve the fidelity of computation.
On top of these designs, a tailored system-level optimization framework can automatically determine the optimal hyperparameters for high performance on the real quantum device.  
Experimental results show that the model searched by our method can achieve a $90\%$ accuracy on an actual quantum device. Compared with that, the accuracy of the existing quantum learning models will drop to roughly $50\%$ for binary classification on an actual quantum device.
More importantly, we show that ST-VQC successfully integrated non-linearity into the quantum learning, which can achieve over $90\% $ prediction on the synthetic dataset with non-linearity.

\section*{Acknowledgment}
This work is supported in part by Mason’s Office of Research Innovation and Economic Impact (ORIEI) and Quantum Science and Engineering Center (QSEC). The research used IBM Quantum resources via Los Alamos National Lab Hub. This material is based upon work supported by the U.S. Department of Energy, Office of Science, National Quantum Information Science Research Centers, Co-design Center for Quantum Advantage (C2QA) under contract number DE-SC0012704, (Basic Energy Sciences, PNNL FWP 76274). In addition, the research used resources from the Oak Ridge Leadership Computing Facility at the Oak Ridge National Laboratory, which is supported by the Office of Science of the U.S. Department of Energy under Contract No. DE-AC05-00OR22725. The Pacific Northwest National Laboratory is operated by Battelle for the U.S. Department of Energy under Contract DE-AC05-76RL01830.

\bibliographystyle{IEEEtran}
\bibliography{bib}
\vspace{12pt}

\end{document}